\documentclass[journal=jpclcd,manuscript=letter]{achemso}
\usepackage{setspace}
\usepackage{amsmath,amssymb}
\usepackage{graphicx}

\usepackage{bm}
\usepackage{multirow}
\usepackage{cases}
\usepackage{afterpage}

\title{Extreme Parametric Sensitivity in the Steady-State Photoisomerization of Two-Dimensional Model Rhodopsin}
\author{Chern Chuang}
\email{chern.chuang@utoronto.ca}
\author{Paul Brumer}
\email{Paul.Brumer@utoronto.ca}
\affiliation{Chemical Physics Theory Group, Department of Chemistry, and Center for Quantum Information and Quantum
Control, University of Toronto, Toronto, Ontario M5S 3H6, Canada}
\begin{document}
\setlength{\fboxrule}{0 pt}

\begin{tocentry}
  \includegraphics{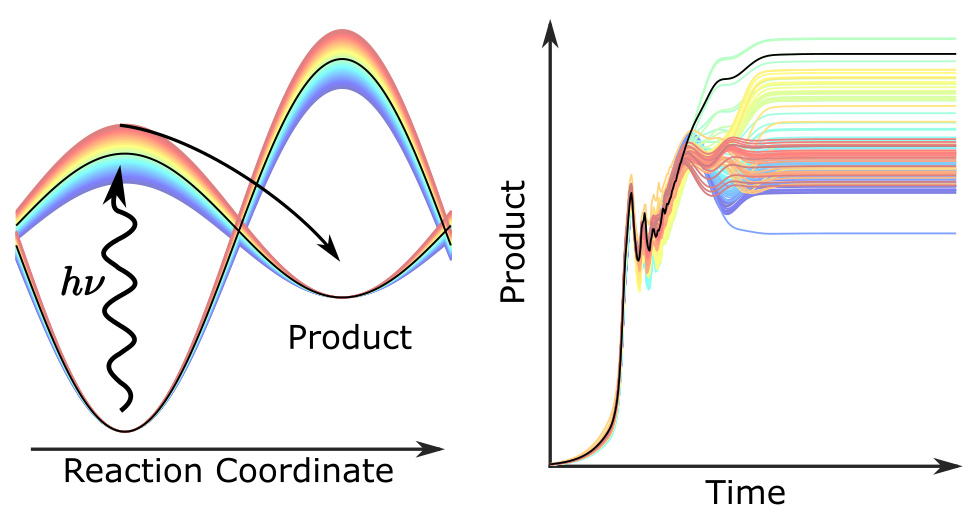}
  For Table of Contents Only
\end{tocentry}

\begin{abstract}
 We computationally studied the photoisomerization reaction of the retinal chromophore in rhodopsin using a two-state two-mode model coupled to thermal baths. Reaction quantum yields at the steady state (10 ps and beyond) were found to be considerably different than their transient values, suggesting a weak correlation between transient and steady-state dynamics in these systems. Significantly,  the steady-state quantum yield was highly sensitive to minute changes in system parameters, while  transient dynamics was nearly unaffected. Correlation of such sensitivity with standard level spacing statistics of the nonadiabatic vibronic system suggests a possible origin in quantum chaos. The significance of this observation of quantum yield parametric sensitivity in biological models of vision has profound conceptual and fundamental implications. 
\end{abstract}

\maketitle
Retinal proteins are a family of important membrane proteins responsible for various photobiological functions in bacteria and higher life forms.\cite{Ottolenghi1982,Sheves2013,SchultenBook2014} Despite the variety of their functions and the host organisms, the general structure of the retinal proteins is consistent, \textit{i.e.} seven transmembrane helices with a retinal chromophore bound to a lysine residue through a protonated Schiff base. Two systems are frequently studied: (1)  rhodopsin that is found in animal retina, responsible for dim light vision, and (2) bacteriorhodopsin found in certain bacteria which functions as a light-driven proton pump. In particular, rhodopsin, with its retinal chromophore initially in an 11-\textit{cis} conformation, undergoes photoisomerization to its all-\textit{trans} counterpart within 200 fs upon pulsed laser excitation.\cite{schoenlein1991,wang1994,polli2010} This has been recently revised to 60 fs. \cite{johnson2017} The short reaction time is attributed to a conical intersection between the ground and excited states, and has been suggested as being responsible for the high photoisomerization quantum yield ($\approx$67\%).\cite{dartnall1968} These observations have prompted the important question of the correlation between quantum coherence and biological function. Our view is clear -- such coherences are induced by the laser pulses, and are irrelevant to nature, which operates with incoherent light.\cite{Tscherbul14,BrumerPerspective2018} Indeed, the natural vision process operates in the steady state regime.

Here, we report the observation of a dramatic dependence of the steady-state quantum yield (QY) on minute changes in system parameters. Our study is based  on a two-state two-mode (2S2M) model coupled to thermal baths  initially proposed by Stock et al.\cite{hahn2000,hahn2002}, and later modified/extended to  account for numerous experimental observations of bovine rhodopsin photoisomerization.\cite{johnson2017,Olivucci2S3M}  The dependence of the biologically important steady-state QY on the system parameters is correlated with  level spacing statistics, a well established measure for quantum chaos in molecular systems. This parametric sensitivity has profound implications for the biology of vision, raising issues as to whether nature's parameters occupy a ``Goldilocks'' region for optimal function.


This letter is arranged as follows. We first introduce the necessary background and details of the theoretical treatment, which is followed by a demonstration of the dramatic parametric sensitivity of the steady-state photoisomerization QY. We then discuss relationships to quantum chaos, the possibility of experimental measurement of such extreme sensitivity, and its implications on generic photochemistry in the condensed phase and biological light-sensing.

{\bf \textit{Models and Method---}} The model for the retinal rhodopsin photoisomerization reaction is simulated numerically using the 2S2M model pioneered by Stock et al.\cite{hahn2000,hahn2002,balzer2005}. This has been shown to be an adequate minimal model for studying the photoisomerization dynamics of retinal chromophores, describing the vibronic structure of a conical intersection.\cite{Olivucci2000PNAS} To account for the environmental effects associated with the protein pocket, the 2S2M system is further coupled to two sets of harmonic bath modes, and the system dynamics is simulated using second order quantum master equations. 

\begin{figure}[htbp]
	\centering
  \includegraphics[height=6cm]{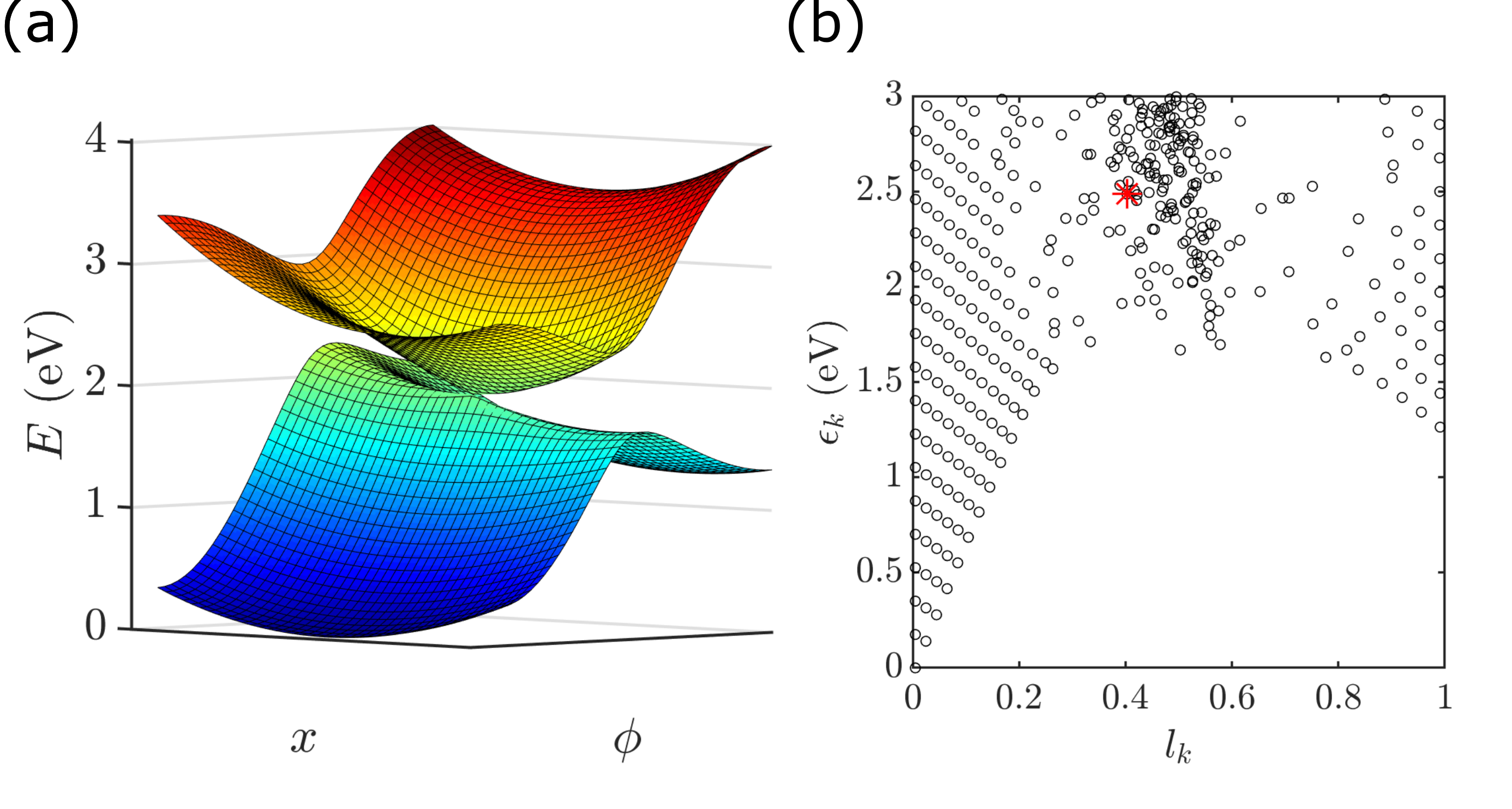}
  \caption{ (a) Adiabatic potential energy surfaces of the 2S2M model. (b) The distribution of the system eigenstates with their energies as the vertical coordinates and $l_k$, defined in Eq.~(\ref{eqn:transness}) as the horizontal coordinates. The state that contributes most to the Franck-Condon state is marked with an asterisk. See text for detailed description.}
\label{fig:2S2M_model}
\end{figure}

A general system-bath Hamiltonian  can be written as
\begin{eqnarray}
H=H_\mathrm{s}+H_\mathrm{b}+H_\mathrm{sb}.
\end{eqnarray}
Here, the system (2S2M) Hamiltonian is comprised of two diabatic electronic states $|0\rangle$ and $|1\rangle$, a reaction coordinate (coupling mode) $\phi$, and a tuning mode $x$.
\begin{eqnarray}
H_\mathrm{s}&=&\sum_{n,n'=0,1}\left[\left(\hat{T}+E_n+(-1)^n\frac{V_n}{2}(1-\cos\phi)+\frac{\omega x^2}{2}+\kappa x\delta_{n,1}\right)\delta_{n,n'}+\lambda x(1-\delta_{n,n'})\right]|n\rangle\langle n'|
\label{eqn:Hsys}
\end{eqnarray}
where $T=-(1/(2m))(\partial^2/\partial\phi^2)+(\omega/2)(\partial^2/\partial x^2)$ is the kinetic energy operator. All  parameters, updated in 2017, are obtained from Ref.[\cite{johnson2017}], and given in Table~\ref{tab:parameters}. We define the range $\phi\in\left[-\pi/2,\pi/2\right)$ to be the \textit{cis-}conformer, and the range $\phi\in\left[\pi/2,3\pi/2\right)$ to be the \textit{trans-}conformer. The eigenstates of $H_\mathrm{s}$ can be expressed in the direct product space of the electronic diabatic states $\{|0\rangle,|1\rangle\}$, the tuning mode $x$, and the reaction coordinate $\phi$.\cite{hahn2000} The adiabatic potential energy surfaces are plotted in Fig.~\ref{fig:2S2M_model}(a). We note that the protein pocket of rhodopsin is such that the retinal chromophore is pre-twisted and chiral, which Eq.~(\ref{eqn:Hsys}) does not account for.\cite{NakamichiOkada2006ACIE}

A generic property of such a Hamiltonian is that the low energy part of its eigenstates is localized in either the \textit{cis-} or the \textit{trans-}wells, while  higher energy  states are more delocalized over the reaction coordinate $\phi$.\cite{tscherbul2014} To better demonstrate this character, of relevance later below, we define the following quantity
\begin{eqnarray}
l_k&=&\frac{1}{2\pi}\int_0^{2\pi}d\phi~|\langle\phi|k\rangle|^2\cdot\frac{(1-\cos\phi)}{2}
\label{eqn:transness}
\end{eqnarray}
where $|k\rangle$ is the $k^{th}$ eigenfunction of the system Hamiltonian $H_\mathrm{s}$, and $|\phi\rangle$ is the eigenfunction of the position operator in the reaction coordinate. This quantity is a non-negative real number with a maximal value of unity. It measures the ``\textit{trans}-ness" of the eigenfunctions: $l_k=0$~$(1)$ corresponds to $|k\rangle$ fully localized in the \textit{cis-} (\textit{trans-}) well. For intermediate values it indicates an averaged position of the eigenstate, localized or delocalized.  Fig.~\ref{fig:2S2M_model}(b) shows the distribution of $l_k$ and eigenstate energies $\epsilon_k=\langle k|H_\mathrm{s}|k\rangle$. Contribution from the \textit{trans} species occurs at most energies, but becomes significant above $\sim1.4$ eV.

\begin{table}[htbp]
\caption{List of parameters, adopted from Ref. [\cite{johnson2017} ]}
\begin{center}
\begin{tabular}{|c|c|}
 \hline
 Parameter & Value (eV) \\
 \hline
 $m^{-1}$ & 2.80$\cdot10^{-3}$ \\
 \hline
 $E_0$ & 0 \\
 \hline
 $E_1$ & 2.58 \\
 \hline
 $V_0$ & 3.56 \\
 \hline
 $V_1$ & 1.19 \\
 \hline
 $w$ & 0.19 \\
 \hline
 $\kappa$ & 0.19 \\
 \hline
 $\lambda$ & 0.19 \\
 \hline
\end{tabular}
\end{center}
\label{tab:parameters}
\end{table}%

\textit{System-Bath Coupling:} To properly account for relaxation effects, the system is coupled to environmental degrees of freedom, approximated by the direct product form
\begin{eqnarray}
H_\mathrm{sb}&=&\sum_{\mathrm{b}} \hat{S}^\mathrm{(b)}\otimes\hat{B}^\mathrm{(b)},\label{eqn:FactorizedSB}
\end{eqnarray}
where $\hat{S}^\mathrm{(b)}$ and $\hat{B}^\mathrm{(b)}$ are operators that  depend only on the system and the bath coordinates, respectively. The superscript $\mathrm{b}$ runs over all system-bath interactions involved. This form of system-bath interaction Hamiltonian is then treated with the standard Markovian-Redfield approximation that leads to a quantum master equation, which in the system eigenbasis reads
\begin{eqnarray}
\frac{\partial}{\partial t}\rho_{ij}&=&-i\omega_{ij}\rho_{ij}+\sum_\mathrm{b}\sum_{kl}\mathcal{R}_{ij,kl}^\mathrm{(b)}\rho_{kl}\\
\mathcal{R}_{ij,kl}^\mathrm{(b)}&=&-\delta_{jl}\sum_r\Gamma_{irrk}^\mathrm{(b)}+\Gamma_{ljik}^\mathrm{(b)}+\left(\Gamma_{ljik}^\mathrm{(b)}\right)^*-\delta_{ik}\sum_r\left(\Gamma_{lrrj}^\mathrm{(b)}\right)^*
\label{eqn:Redfield}
\end{eqnarray}
where $\omega_{ij}$ is the energy difference between the system eigenstates $i$ and $j$, and  $\hbar=1$. The damping tensors $\Gamma_{ijkl}^\mathrm{(b)}$ are given by
\begin{eqnarray}
\Gamma_{ijkl}^\mathrm{(b)}=S_{ij}^\mathrm{(b)}S_{kl}^\mathrm{(b)}J^\mathrm{(b)}(\omega_{kl})\bar{n}_\mathrm{BE}(\omega_{kl},\beta),
\label{eqn:Gamma}
\end{eqnarray}
where $S_{ij}^\mathrm{(b)}=\langle i|\hat{S}^\mathrm{(b)}|j\rangle$, $\bar{n}_\mathrm{BE}(\omega,\beta)=1/(e^{\beta\omega}-1)$ is the Bose-Einstein distribution at inverse temperature $\beta=1/k_\mathrm{B}T$ with the Boltzmann constant $k_\mathrm{B}$.

The computational cost of propagating the Redfield master equation  [Eq.~(\ref{eqn:Redfield})] can be formidable given that one typically needs up to a thousand lowest energy eigenstates, enough to converge the Franck-Condon transition from the ground to excited state. Significantly, this is made more difficult by the fact that we need to examine the long time steady-state properties, which requires propagating the dynamics up to a nanosecond and beyond. To this end, one can invoke the Bloch-secular approximation which decouples the population dynamics from that of the coherences.\cite{balzer2005} We compare the results from the full nonsecular treatment to those of the secular approximation in the next section. (For issues of secular v.s. nonsecular master equations, see Ref.[\cite{Dodin2018PRA}].)

The phonon bath is described within the harmonic approximation, $H_\mathrm{ph}=\sum_k\omega_k\cdot b^\dagger_kb_k$, where the system-phonon interaction can be further separated into two components:
\begin{eqnarray}
H_\mathrm{s-ph}&=&H_{\mathrm{s-}x}+H_{\mathrm{s}-\phi}\\
H_{\mathrm{s-}x}&=&|1\rangle\langle1|~x\cdot\sum_kg_{k,x}(b_{k,x}^\dagger+b_{k,x})\\
H_{\mathrm{s-}\phi}&=&|1\rangle\langle1|(1-\cos\phi)\cdot\sum_kg_{k,\phi}(b_{k,\phi}^\dagger+b_{k,\phi})
\label{eqn:sys-phonon}
\end{eqnarray}
where $g_{k,x}$ and $g_{k,\phi}$ are the coupling strengths of the phonon mode $k$. It is customary to represent the summation over the phonon coupling strengths $g_k$ as a continuous spectra density $J(\omega)$, taken here in accord with the work of Stock et al.,\cite{hahn2002,balzer2005} to be an Ohmic bath with exponential cut-off, $J(\omega)=\eta\omega e^{-\omega/\omega_c}$. Couplings to these baths represent the fluctuations and dissipation of the system dynamics along the modes $x$ and $\phi$. 

In the simulation of the retinal photoisomerization, it is customary to choose the initial condition of the dynamics to be the Franck-Condon state from the ground state located in the \textit{cis-}well. \cite{hahn2000,johnson2017,sala2018} On the other hand, the dynamics initiated from a state similar to the Franck-Condon state but with all coherence in the energy eigenbasis removed, referred to as the Franck-Condon mixed state, mimics the effect of an incoherent light source such as sunlight.\cite{tscherbul2014} We focus on the dynamics initiated with the Franck-Condon states and compare the effect of incoherent Franck-Condon mixed initial state in the Supporting Information. Similar parametric sensitivity was also observed in the non-equilibrium steady state case, where the retinal system is simultaneously connected to a photon bath describing the incoherent sunlight and a phonon bath describing the protein environment. (See Supporting Information.)

{\bf \textit{Variation of System Parameters---}} As noted above, the main result of this work is our observation of the extreme sensitivity of the steady-state photoisomerization QY to parameter variation. To demonstrate the extreme sensitivity, consider the results of varying two representative parameters, as manifest in the computed steady-state QY. First, we vary the inverse mass $m^{-1}$ in the kinetic energy term of the reaction coordinate. In the 2S2M model, $m^{-1}$ determines the characteristic frequency of the reaction coordinate in the harmonic regime, $\omega_\phi=\sqrt{k/m}$ and $k=V_0/2$. It is worth noting that in the original model of Hahn and Stock\cite{hahn2000} $m^{-1}=4.84\cdot10^{-4}$ (eV),  nearly six-fold smaller than the value recently suggested by Johnson et al.\cite{johnson2017}, introduced to reflect the shortening of the rise time of the $trans-$photoproduct observed in pump-probe and 2D photon echo measurements with better temporal resolution. Here we consider $\pm5\%$ variation of $m^{-1}$ about the value proposed by Johnson et al. Note that this corresponds a small change, \textit{i.e.} $~\pm2.5\%$, of the harmonic frequency along the reaction coordinate.

We have also examined the effect of varying the optical gap between the two diabatic electronic states, namely the parameter $E_1$. It is estimated that the inhomogeneous contribution to the broad absorption band of retinal rhodopsin is $\sim$1000 cm$^{-1}$\cite{hahn2000CP}, $\sim5\%$ of the optical gap, and which is the range over which we vary the optical gap. To avoid the complication associated with changing the energy storage ($E_1-V_1$) and the optical gap of the $trans-$product ($V_0-E_1+V_1$) when varying $E_1$ we simultaneously change $V_1$ to keep $E_1+V_1$ constant.

\begin{figure}[htbp]
	\centering
  \includegraphics[height=7cm]{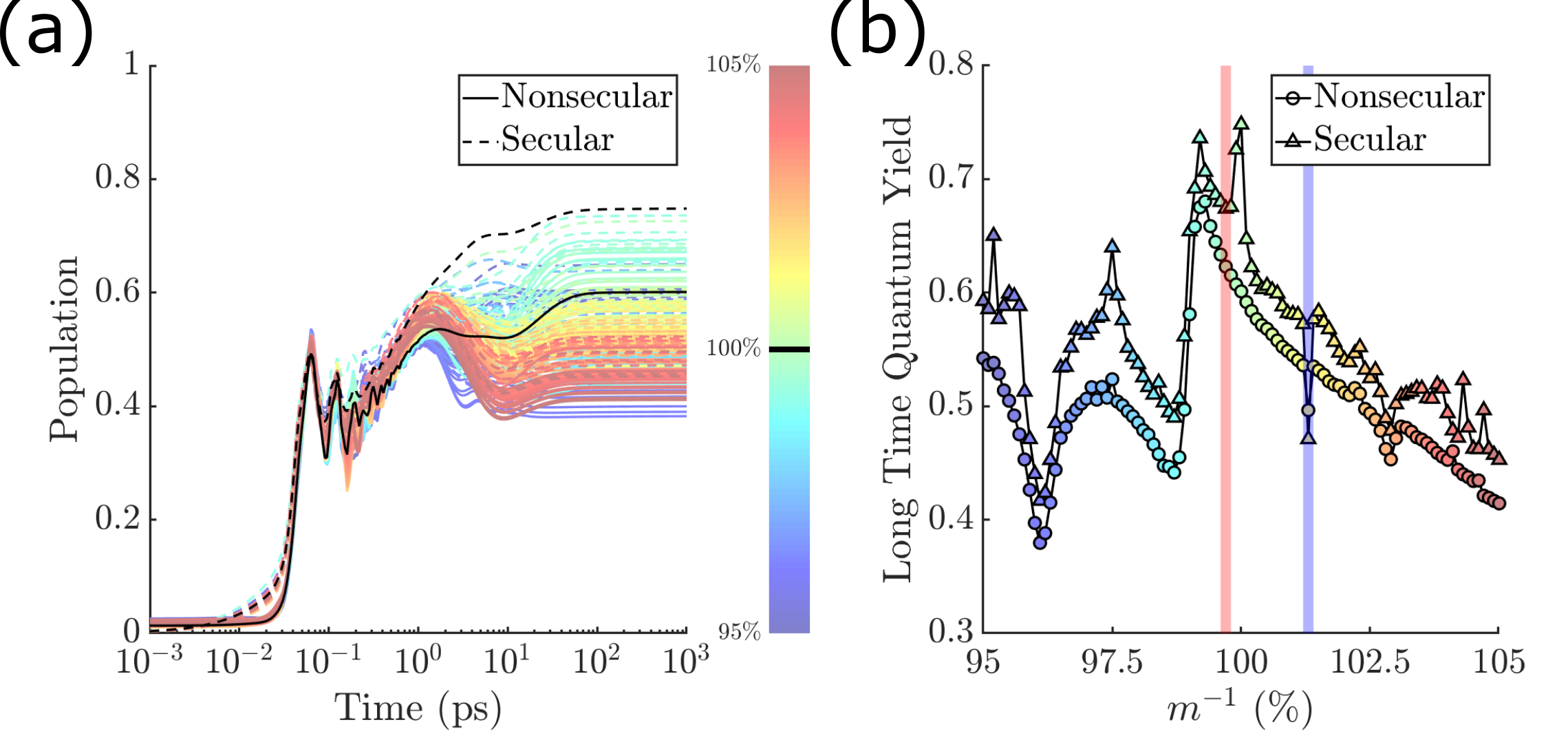}
  \caption{(a) Time dependent $trans$-photoproduct population. The full nonsecular dynamics is shown as solid lines and those of the Bloch-secular are shown with dashed lines. The data corresponding to the original set of parameters given in Table \ref{tab:parameters} are shown in black. Data with varied $m^{-1}$ are color coded as indicated by the color bar. (b) QY recorded at 10 ns. The blue and red shaded area correspond to the connectivity graphs shown in Fig.~\ref{fig:NNSD}, discussed later below. The bath parameters are adopted from Ref.[\cite{balzer2005}].}
\label{fig:QYpDep}
\end{figure}

\textit{Variation of $m^{-1}$:} Fig.~\ref{fig:QYpDep}(a) shows the time-dependent $trans$-ground state population of the photoisomerization dynamics, which gives the ``time-dependent quantum yield", defined as the population of diabatic electronic state $|1\rangle$ in the $trans-$region ($\pi/2\le\phi<3\pi/2$).  The dynamics of the original model with parameters given in Table.~\ref{tab:parameters} is shown in black, and those with modified $m^{-1}$ parameters are color-coded. The bath temperature is kept at 0K to avoid the complication of thermally activated isomerization. Test computations with the bath at room temperature were also carried out to ensure that the parameter dependences reported below remained qualitatively unchanged. 

Note first that there are signs of coherent beatings at times less than 1 ps reflecting the fact that the population is oscillating between the reactant and the product wells, as the population in the \textit{cis-}well mirrors that of the \textit{trans-} (data not shown). The photoproduct appears on the time scale of 100 fs alongside the beatings, in general agreement with ultrafast pulsed laser experiments,\cite{schoenlein1991,wang1994,polli2010,johnson2017} signaling coherent wavepacket dynamics along the direction of the reaction coordinate $\phi$. While the beatings disappear on the ps time scale, the QY continues to fluctuate on a far longer time scale, up to 100 ps. This is consistent with the experimental observations of a closely related system, the photoisomerization of bacteriorhodopsin, which shows a similar time-dependent photoproduct signal (up to hundreds of ps) after the pump pulse.\cite{prokhorenko2006} Computationally, the photoproduct signal has a non-monotonic time dependence, a steep rise on the time scale of 1 ps, followed by a smooth decay until 20 ps, reaching its stationary value at longer times.

Here we are varying the particular system parameter $m^{-1}$ within the range of $\pm5\%$ with respect to the original value. The transient dynamics, up to $\sim$0.5 ps, show limited $m^{-1}$ dependence. However, most significantly, the QY at longer times (and hence not observable in transient dynamics reported in most pulsed laser experiments) depends very sensitively on $m^{-1}$, spanning a wide range between $35\%$ and $75\%$. This is clearly seen in the QY recorded at 10 ns as a function of $m^{-1}$ shown in Fig.~\ref{fig:QYpDep}(b). Note also that Fig.~\ref{fig:QYpDep} shows both secular and nonsecular simulations. In general, these two sets of results agree qualitatively with QY sensitive to the value of $m^{-1}$. However, it is clear that the steady-state QY depends on the level of treatment for the system-bath interaction.


\begin{figure}[htbp]
	\centering
  \includegraphics[height=7cm]{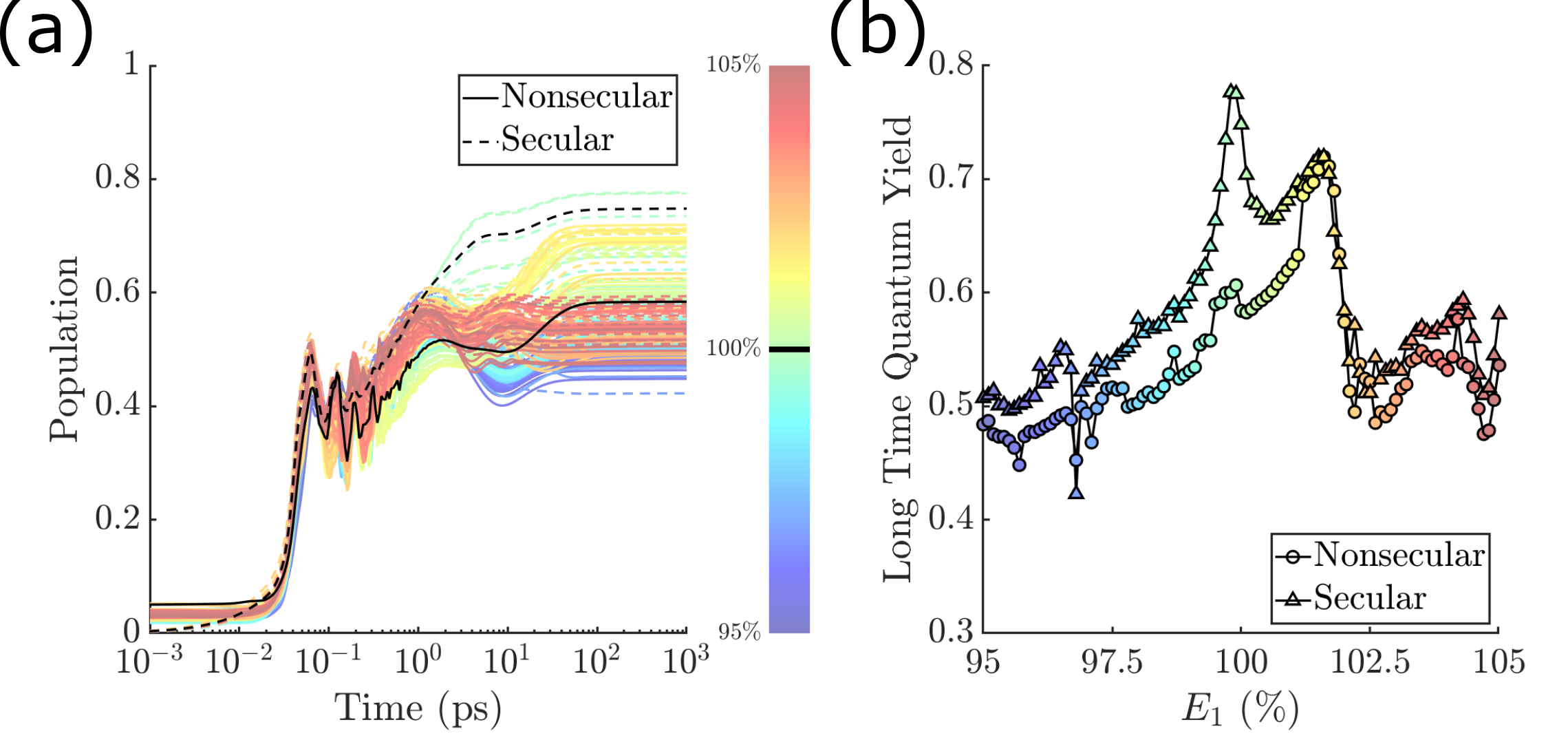}
  \caption{As in Fig.~\ref{fig:QYpDep}, but with $E_1$ is changed instead of $m^{-1}$.}
\label{fig:QYEcDep}
\end{figure}

\textit{Variation of The Optical Gap:} To corroborate the above result and ensure that the parameter $m^{-1}$ is not particularly special with regard to the sensitivity of steady-state QY, we also examined the system dependence on the value of the optical gap. This quantity is known to be  inhomogeneously broadened in natural systems and easier to measure in concert with the photoisomerization reaction itself. In our case, this amounts to changing the parameter $E_1$ in the 2S2M model as described in the Models and Method section. The results are shown in Fig.~\ref{fig:QYEcDep}, and are  similar to those found in variation of $m^{-1}$, \textit{i.e.} large variations in QY as a function of small changes in the optical gap.

We also examined other parameters such as $\omega$ (the frequency of the harmonic, second mode in the 2S2M model) and $\lambda$ (the magnitude of the nonadiabatic coupling, similar to that studied in Ref.[\cite{Takatsuka2001PRE}]) and, moreover, two-state-one-mode and two-state-two-mode models with avoided crossings instead of a conical intersection (parameters given in Ref.[\cite{balzer2005}]). In light of recent computational study,\cite{Olivucci2S3M,Marsili2020JCTC} we simulated and found similar pararmetric sensitivity in a two-state-three-mode model as well. In addition to the above mentioned nonadiabatic, two-surface models, we consider a single-surface double well (1S2M) model as well that is more suitable for describing the adiabatic regime (strong nonadiabatic coupling) of the isomerization reaction. Details are provided in the Supporting Information. In all cases similar extreme sensitivities of steady-state QY are found with respect to variations of system parameters. This suggests that the phenomenon may be an inherent property of the multi-dimensional vibronic/vibrational systems.\cite{Heller1990,LeitnerCederbaum1996} 

{\bf \textit{System Eigenstate Statistics and Relaxation Pathways---}} We conjecture that the parametric sensitivity of the steady-state QY results from the sensitivity of chaotic system dynamics that moves between the two wells and eventually settles into the product well by relaxation induced by the system-bath coupling. As mentioned in the introduction, previous studies on similar systems in the quantum chaos literature  focus mainly  on the statistical properties of their eigenstates\cite{Heller1990} or short-time trajectory-based simulations in the absence of thermal relaxation.\cite{Cederbaum1983,SantoroOlivucci2007} The system part in the current study, Eq.~(\ref{eqn:Hsys}), falls under the same category of nonadiabatic vibronic Hamiltonians.\cite{Heller1990,Takatsuka2001PRE} It is thus expected, if the open system character is not significant, that similar quantum chaos characteristics would exist in this system as well.  Fig.~\ref{fig:2S2M_model}(b) shows the distribution of eigenstates according to their energies and $l_k$, a measure of the average position along the reaction coordinate. It is clear that in the low energy and deeply $cis$/$trans$ regions ($l_k\rightarrow0$ or $l_k\rightarrow1$), the distribution of eigenstates forms a regular grid point lattice. This implies that these states can be written as direct products of the eigenstates of the two modes, $|k\rangle\propto|n_\phi\rangle\otimes|n_x\rangle$, where $n_\phi$ and $n_x$ are the quantum numbers of the respective modes. On the other hand, for states with higher energies and $l_k$ closer to 0.5, the grid pattern disappears and the characters of the two diabatic electronic states and the two modes mix.    

The nearest-neighbor spacing distribution (NNSD) of energy eigenstates is a commonly utilized measure to characterize possible chaotic nature of an isolated system. \cite{Cederbaum1983,Takatsuka2001PRE} Specifically, level repulsion tends to imply chaos, most clearly manifest in a Wigner distribution of adjacent energy level spacings. Here, the procedure described by Haller et al. is used  to account for the smoothly varying part of the density of states.\cite{Cederbaum1983} The ``unfolded" energy levels are defined as
\begin{eqnarray}
\tilde{\epsilon}_{i+1}=\tilde{\epsilon}_{i}+(2k+1)\frac{\epsilon_{i+1}-\epsilon_i}{\epsilon_{j_2+1}-\epsilon_{j_1}}
\end{eqnarray}
where $\epsilon_i$ are the energies of the eigenstates of $H_\mathrm{s}$ with the state index $i$ that runs from $1$ to $n$, $k$ is a small positive integer defining the locality of the unfolding, $j_1=\mathrm{max}(1,i-k)$ and $j_2=\mathrm{min}(n-1,i+k)$.  NNSD is the distribution function of $S=\tilde{\epsilon}_{i+1}-\tilde{\epsilon}_i$. A non-monotonic NNSD signals quantum chaotic behavior. Most notably, by assuming linear repulsion between adjacent levels one arrives at the Wigner distribution for chaotic systems:
\begin{eqnarray}
P_\mathrm{w}(S)=\frac{\pi S}{2D^2}e^{-\frac{\pi S^2}{4D^2}}
\label{eqn:Wigner}
\end{eqnarray}
where $D$ is the mean of the distribution.

\begin{figure}[htbp]
\begin{minipage}{\textwidth}
	\centering
  \includegraphics[height=5.5cm]{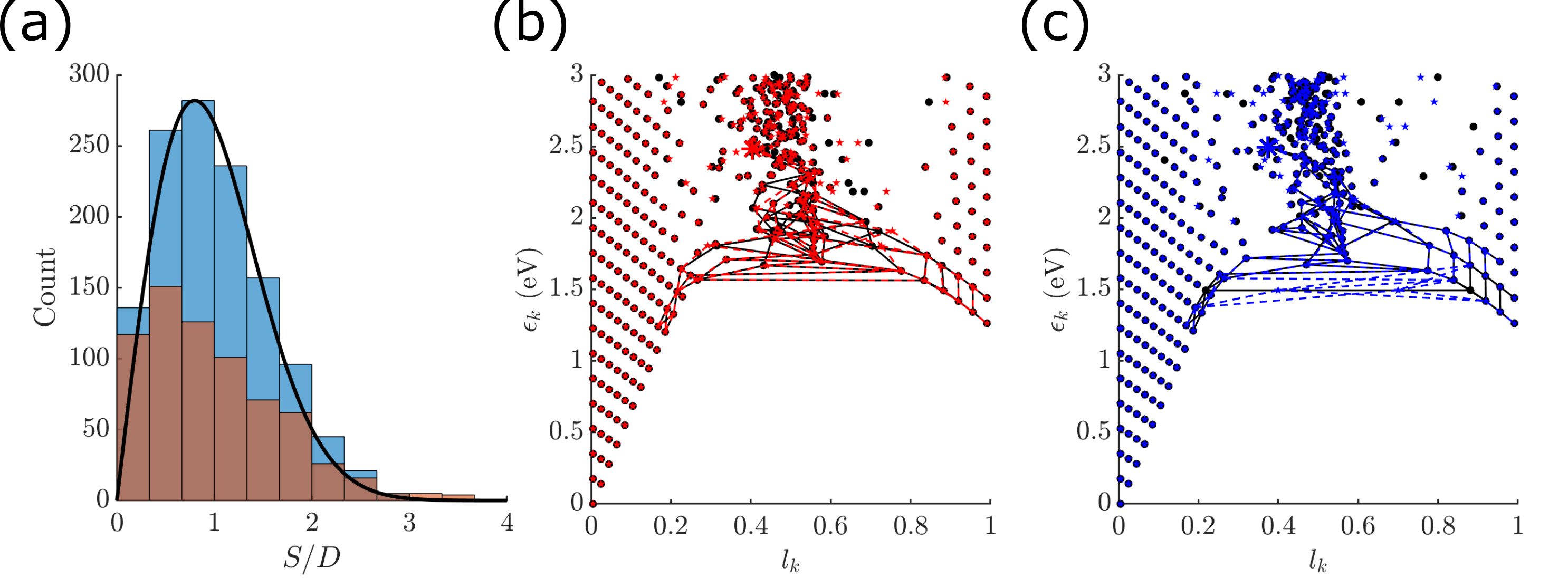}
  \caption[Caption for LOF]{(a)  The nearest-neighbor energy level spacing distribution of the model system studied, using parameters in Table~\ref{tab:parameters}. The blue bars (higher) are the 1240 eigenstates taken from the energy range 4 to 6.4 eV, and the brown ones (lower) are the 686 eigenstates with energy below 4 eV. The local mean spacing parameter is set to $k=2$. The average spacing of the unfolded levels is $D=1.00$ meV. The curve is the Wigner distribution, given in Eq.~(\ref{eqn:Wigner}).\footnote{The adoption of the higher energy range in Fig. 4(a) is for the purpose of emphasizing the contribution from the chaotic regime. In all other instances in the paper the photoisomerization dynamics is well converged by including energy eigenstates below 3 eV.} (b) The connectivity tree graph of the systems with $m^{-1}$ values 99.70\% and 99.80\% of that in Table.~\ref{tab:parameters} [red shaded region in Fig.~\ref{fig:QYpDep}(b)], rooted at the brightest state (asterisk). (c) The tree graph of the systems with $m^{-1}$ 101.3\% and 101.4\% of that in Table.~\ref{tab:parameters} [blue shaded region in Fig.~\ref{fig:QYpDep}(b)].}
\label{fig:NNSD}
\end{minipage}
\end{figure}

The NNSD for our system is shown in Fig.~\ref{fig:NNSD}(a). It is clear that the distribution is non-monotonic and near Wignerian and hence chaotic. This is especially true for the higher energy part of the eigenspectrum (blue), in comparison to the lower energy counterpart (brown). 

Quantum chaos in the context of vibronic eigenspectrum of small molecules/ions in the gas phase has been well established. However, the situation is less clear with systems, such as ours, coupled to a dissipative environment. As one insightful route, we analyze the relaxation pathways from the  brightest state (the eigenstate contributing the most to the Franck-Condon state) by examining the Bloch-secular rate matrix elements among the eigenstates. 


In Fig.~\ref{fig:NNSD}(b) and (c) we show the tree graphs representing the connectivity matrix\cite{BiggsGraphTheoryBook} of the relaxation pathways among system energy eigenstates calculated using two pairs of $m^{-1}$ values differed by only $0.1\%$, corresponding to the red and blue shaded regions shown in Fig.~\ref{fig:QYpDep}(b), while keeping all other parameters fixed. Each graph shows the eigenstate distributions [as in Fig.~\ref{fig:2S2M_model}(b)] of the two system Hamiltonians with slightly different $m^{-1}$ values, using different markers (black filled circles and colored stars). As shown in Fig.~\ref{fig:QYpDep}(b), the red region corresponds to a parameter regime that shows smooth changes of steady-state QY with changing $m^{-1}$ values, whereas the blue region shows dramatic changes. 
To simplify the graphs we only connect each higher energy state $i$ to the two lower energy states $j$ with the largest scattering rates $\sum_\mathrm{b}\mathcal{R}^\mathrm{(b)}_{jjii}$ in Eq.~(\ref{eqn:Redfield}). 
The connectivity graphs are rooted at the brightest state of the corresponding systems, defining tree graphs of degree two. 

In both pairs of the tree graphs it can be seen that very small changes in $m^{-1}$ leaves the approximated direct product structure for states close to the \textit{cis}- ($l_k\approx0$) and the \textit{trans}-wells ($l_k\approx1$) nearly unaffected. That is, a 0.1\% change of $m^{-1}$ corresponds to a $\sqrt{10^{-3}}\approx3\%$ change in the harmonic frequency along the reaction coordinate (the vertical spacing between the lowest and the second lowest symbols), generally not perceivable by the naked eye. By contrast, a significant portion of the states with $l_k$ close to 0.5 shows a noticeable horizontal shift, signifying the existence of level repulsion discussed above. In addition, a change in the eigenstate structure induces a dramatic change in the  steady-state QY only if the shifted states are involved in the dominant relaxation pathways. This is so in the case shown in Fig.~\ref{fig:NNSD}(c) in contrast to that in Fig.~\ref{fig:NNSD}(b), in agreement with the respective behavior of the steady-state QY shown in Fig.~\ref{fig:QYpDep}(b). These observations complement the parametric sensitivity of long-time dynamics displayed in Figs.~\ref{fig:QYpDep} and \ref{fig:QYEcDep}, supporting the view that isolated system chaos propagates through the relaxation dynamics in the open system.



{\bf \textit{Possible experimental verification---}}
We note the recent work from Schnedermann et al., where rhodopsins reconstituted with isotope-substituted retinal chromophores show similar short-time dynamics, but very different steady-state QY.\cite{schnedermann2018} This observation supports the view that a shorter rise time (the time it takes for the first peak of photoproduct signal to appear) corresponds to faster reaction, but not necessarily to higher reaction QY. That is, there is a disconnect between transient and steady state behavior. However, changes with isotope substitution are not directly comparable to the sensitivity of steady-state QY seen in Figs.~\ref{fig:QYpDep} and \ref{fig:QYEcDep}. First, the isotope labelings in these experiments constitute much stronger perturbations to the system than those studied here. Second, the experimental observations are typically, if not exclusively, made on large ensembles of molecules. 
Within the ensemble, one expects significant inhomogeneity among the parameters, which would wash out the kind of parametric sensitivity observed in our computations.

For the real retinal system in an opsin protein pocket, the experimental situation is expected to be much more complicated than we have modeled above. Indeed, the established methods of measuring steady-state QY typically involve macroscopic amounts of the sample being photoisomerized (for example loss of $\approx10\%$ of the optical density of the reactant over the course of a few minutes\cite{kim2001,Kukura2012JACS}). This being the case, and as noted above, it is likely that the sensitivity observed here is washed out by the sheer number of retinal molecules in a sample ($\approx10^{20}$), each possessing a different conformation. Hence, experimental confirmation of the sensitivity of QY will require measurements of small ensembles or even of single molecules.  Provided that such sensitivity can be verified experimentally for retinal systems, it may well be the case that one can find similar phenomena in other light-induced condensed phase curve-crossing systems.


{\bf \textit{Open Theoretical Challenges---}} Additional investigation is needed to clarify the relationship between quantum chaos and parametric sensitivity of meta-stable state populations in an open system under environmental dissipation. Whereas we continue to examine a number of possible measures of quantum chaos for open systems proposed in the literature, such as the Loschmidt echos and the ratio of eigenvalue differences of the Liouville operator,\cite{GrobeHaakeSommers1988PRL,Prosen2019PRL,Prosen2020PRX,JalabertPastawski2001PRL,CucchiettiZurek2003PRL} results thus far are inconclusive. (These results are available upon reasonable request.) For example, they suggest a need for a criteria that applies to open systems whose isolated system component displays both integrable and chaotic energy level regions. Further, it is worth noting that our treatment of system-bath interactions is restricted to a Markovian and perturbative (fast bath) limit, with other environmental effects included as static disorder (slow bath). In doing so, we do not detect QY sensitivity to the parameters of the system-bath coupling term or to the bath itself. In the future it would be interesting to examine non-pertubative system-bath effects using more sophisticated methods that also account for the intermediate regime in which the system and the bath time scales overlap.\cite{DomckeZhao2016,sala2018} 


Furthermore, while admittedly premature, it is not difficult to suggest a connection between this result and its possible significant biological implications. That is, does a biological light-sensing apparatus benefit from the high sensitivity of the underlying curve-crossing mechanism to parameter variation? Do the system parameters have to be in a specific (``Goldilocks") regime of parameter space? Has natural selection tuned these parameters? These and related questions warrant further study. What is the key here is the highly significant possibility of quantum chaotic contributions and parametric sensitivity in biological processes and its role in tuning biological function.

{\bf \textit{Conclusion---}} The photoisomerization dynamics of retinal chromophore in rhodopsin was studied using a two-state-two-mode model coupled to dissipative phonon baths. Assuming the Franck-Condon state as the initial state, the photoproduct (\textit{trans-}conformer) populations first show a rapid rise with strong oscillatory features (quantum beats) on the 100 fs time scale. The oscillations then fade out and are replaced by a smooth time-dependent profile leading to the stationary values on longer time scales. Time-dependent behavior of this type agrees with available experimental observations of the closely related system bacteriorhodopsin, suggesting that simple pictures that correlate short time transient spectral features with steady-state quantum yield are suspect.

Most significantly, we find that the steady-state quantum yield is highly sensitive to slight changes in system parameters, while the transient dynamics remains almost unaffected. Such sensitivity is also observed by changing numerous other parameters of the model system, suggesting that it is an inherent feature of the nonadiabatic vibronic system studied. Further, possible experimental measurements of such sensitivity are discussed. The implications of these results are profound in biological light-sensing systems , and possibly in condensed phase photochemistry.

\textit{Acknowledgement---} Support under AFOSR grant FA9550-19-1-0267 is gratefully acknowledged, as are discussions with Professor Jennifer Ogilvie, University of Michigan and Professor Arjendu Pattanayak, Carleton College.

\section*{Supporting Information} Parametric sensitivity of steady-state photoisomerization observed with incoherent initial conditions and incoherent excitations, parametric sensitivity and nonsensitivity in single surface models.

\newcommand{\noopsort}[1]{} \newcommand{\printfirst}[2]{#1}
  \newcommand{\singleletter}[1]{#1} \newcommand{\switchargs}[2]{#2#1}
\providecommand{\latin}[1]{#1}
\makeatletter
\providecommand{\doi}
  {\begingroup\let\do\@makeother\dospecials
  \catcode`\{=1 \catcode`\}=2 \doi@aux}
\providecommand{\doi@aux}[1]{\endgroup\texttt{#1}}
\makeatother
\providecommand*\mcitethebibliography{\thebibliography}
\csname @ifundefined\endcsname{endmcitethebibliography}
  {\let\endmcitethebibliography\endthebibliography}{}


\begin{mcitethebibliography}{38}
\providecommand*\natexlab[1]{#1}
\providecommand*\mciteSetBstSublistMode[1]{}
\providecommand*\mciteSetBstMaxWidthForm[2]{}
\providecommand*\mciteBstWouldAddEndPuncttrue
  {\def\EndOfBibitem{\unskip.}}
\providecommand*\mciteBstWouldAddEndPunctfalse
  {\let\EndOfBibitem\relax}
\providecommand*\mciteSetBstMidEndSepPunct[3]{}
\providecommand*\mciteSetBstSublistLabelBeginEnd[3]{}
\providecommand*\EndOfBibitem{}
\mciteSetBstSublistMode{f}
\mciteSetBstMaxWidthForm{subitem}{(\alph{mcitesubitemcount})}
\mciteSetBstSublistLabelBeginEnd
  {\mcitemaxwidthsubitemform\space}
  {\relax}
  {\relax}

\bibitem[Ottolenghi(1982)]{Ottolenghi1982}
Ottolenghi,~M. Molecular Aspects of the Photocycles of Rhodopsin and
  Bacteriorhodopsin: A Comparative Overview. \emph{Meth. Enzymol.}
  \textbf{1982}, \emph{88}, 470--491\relax
\mciteBstWouldAddEndPuncttrue
\mciteSetBstMidEndSepPunct{\mcitedefaultmidpunct}
{\mcitedefaultendpunct}{\mcitedefaultseppunct}\relax
\EndOfBibitem
\bibitem[Wand \latin{et~al.}(2013)Wand, Gdor, Zhu, Sheves, and
  Ruhman]{Sheves2013}
Wand,~A.; Gdor,~I.; Zhu,~J.; Sheves,~M.; Ruhman,~S. Shedding New Light on
  Retinal Protein Photochemistry. \emph{Ann. Rev. Phys. Chem.} \textbf{2013},
  \emph{64}, 437--458\relax
\mciteBstWouldAddEndPuncttrue
\mciteSetBstMidEndSepPunct{\mcitedefaultmidpunct}
{\mcitedefaultendpunct}{\mcitedefaultseppunct}\relax
\EndOfBibitem
\bibitem[Schulten(2014)]{SchultenBook2014}
Schulten,~K. In \emph{Quantum Effects in Biology}; Mohseni,~M., Omar,~Y.,
  Engel,~G.~S., Plenio,~M.~B., Eds.; Cambridge University Press: Cambridge, UK,
  2014; Chapter 11, pp 237--263\relax
\mciteBstWouldAddEndPuncttrue
\mciteSetBstMidEndSepPunct{\mcitedefaultmidpunct}
{\mcitedefaultendpunct}{\mcitedefaultseppunct}\relax
\EndOfBibitem
\bibitem[Schoenlein \latin{et~al.}(1991)Schoenlein, Peteanu, Mathies, and
  Shank]{schoenlein1991}
Schoenlein,~R.; Peteanu,~L.; Mathies,~R.; Shank,~C. The First Step in Vision:
  Femtosecond Isomerization of Rhodopsin. \emph{Science} \textbf{1991},
  \emph{254}, 412--415\relax
\mciteBstWouldAddEndPuncttrue
\mciteSetBstMidEndSepPunct{\mcitedefaultmidpunct}
{\mcitedefaultendpunct}{\mcitedefaultseppunct}\relax
\EndOfBibitem
\bibitem[Wang \latin{et~al.}(1994)Wang, Schoenlein, Peteanu, Mathies, and
  Shank]{wang1994}
Wang,~Q.; Schoenlein,~R.~W.; Peteanu,~L.~A.; Mathies,~R.~A.; Shank,~C.~V.
  Vibrationally Coherent Photochemistry in the Femtosecond Primary Event of
  Vision. \emph{Science} \textbf{1994}, \emph{266}, 422--424\relax
\mciteBstWouldAddEndPuncttrue
\mciteSetBstMidEndSepPunct{\mcitedefaultmidpunct}
{\mcitedefaultendpunct}{\mcitedefaultseppunct}\relax
\EndOfBibitem
\bibitem[Polli \latin{et~al.}(2010)Polli, Alto{\`e}, Weingart, Spillane,
  Manzoni, Brida, Tomasello, Orlandi, Kukura, Mathies, \latin{et~al.}
  others]{polli2010}
Polli,~D.; Alto{\`e},~P.; Weingart,~O.; Spillane,~K.~M.; Manzoni,~C.;
  Brida,~D.; Tomasello,~G.; Orlandi,~G.; Kukura,~P.; Mathies,~R.~A.
  \latin{et~al.}  Conical Intersection Dynamics of the Primary
  Photoisomerization Event in Vision. \emph{Nature} \textbf{2010}, \emph{467},
  440\relax
\mciteBstWouldAddEndPuncttrue
\mciteSetBstMidEndSepPunct{\mcitedefaultmidpunct}
{\mcitedefaultendpunct}{\mcitedefaultseppunct}\relax
\EndOfBibitem
\bibitem[Johnson \latin{et~al.}(2017)Johnson, Farag, Halpin, Morizumi,
  Prokhorenko, Knoester, Jansen, Ernst, and Miller]{johnson2017}
Johnson,~P.~J.; Farag,~M.~H.; Halpin,~A.; Morizumi,~T.; Prokhorenko,~V.~I.;
  Knoester,~J.; Jansen,~T.~L.; Ernst,~O.~P.; Miller,~R.~D. The Primary
  Photochemistry of Vision Occurs at the Molecular Speed Limit. \emph{J. Phys.
  Chem. B} \textbf{2017}, \emph{121}, 4040--4047\relax
\mciteBstWouldAddEndPuncttrue
\mciteSetBstMidEndSepPunct{\mcitedefaultmidpunct}
{\mcitedefaultendpunct}{\mcitedefaultseppunct}\relax
\EndOfBibitem
\bibitem[Dartnall(1968)]{dartnall1968}
Dartnall,~H. The Photosensitivities of Visual Pigments in the Presence of
  Hydroxylamine. \emph{Vision Res.} \textbf{1968}, \emph{8}, 339--358\relax
\mciteBstWouldAddEndPuncttrue
\mciteSetBstMidEndSepPunct{\mcitedefaultmidpunct}
{\mcitedefaultendpunct}{\mcitedefaultseppunct}\relax
\EndOfBibitem
\bibitem[Tscherbul and Brumer(2014)Tscherbul, and Brumer]{Tscherbul14}
Tscherbul,~T.~V.; Brumer,~P. Long-Lived Quasistationary Coherences in a V-type
  System Driven by Incoherent Light. \emph{Phys. Rev. Lett.} \textbf{2014},
  \emph{113}, 113601\relax
\mciteBstWouldAddEndPuncttrue
\mciteSetBstMidEndSepPunct{\mcitedefaultmidpunct}
{\mcitedefaultendpunct}{\mcitedefaultseppunct}\relax
\EndOfBibitem
\bibitem[Brumer(2018)]{BrumerPerspective2018}
Brumer,~P. Shedding (Incoherent) Light on Quantum Effects in Light-Induced
  Biological Processes. \emph{J. Phys. Chem. Lett.} \textbf{2018}, \emph{9},
  2946--2955\relax
\mciteBstWouldAddEndPuncttrue
\mciteSetBstMidEndSepPunct{\mcitedefaultmidpunct}
{\mcitedefaultendpunct}{\mcitedefaultseppunct}\relax
\EndOfBibitem
\bibitem[Hahn and Stock(2000)Hahn, and Stock]{hahn2000}
Hahn,~S.; Stock,~G. Quantum-Mechanical Modeling of the Femtosecond
  Isomerization in Rhodopsin. \emph{J. Phys. Chem. B} \textbf{2000},
  \emph{104}, 1146--1149\relax
\mciteBstWouldAddEndPuncttrue
\mciteSetBstMidEndSepPunct{\mcitedefaultmidpunct}
{\mcitedefaultendpunct}{\mcitedefaultseppunct}\relax
\EndOfBibitem
\bibitem[Hahn and Stock(2002)Hahn, and Stock]{hahn2002}
Hahn,~S.; Stock,~G. Ultrafast cis-trans Photoswitching: A Model Study. \emph{J.
  Chem. Phys.} \textbf{2002}, \emph{116}, 1085--1091\relax
\mciteBstWouldAddEndPuncttrue
\mciteSetBstMidEndSepPunct{\mcitedefaultmidpunct}
{\mcitedefaultendpunct}{\mcitedefaultseppunct}\relax
\EndOfBibitem
\bibitem[Marsili \latin{et~al.}(2019)Marsili, Farag, Yang, De~Vico, and
  Olivucci]{Olivucci2S3M}
Marsili,~E.; Farag,~M.~H.; Yang,~X.; De~Vico,~L.; Olivucci,~M. Two-State,
  Three-Mode Parametrization of the Force Field of a Retinal Chromophore Model.
  \emph{J. Phys. Chem. A} \textbf{2019}, \emph{123}, 1710--1719\relax
\mciteBstWouldAddEndPuncttrue
\mciteSetBstMidEndSepPunct{\mcitedefaultmidpunct}
{\mcitedefaultendpunct}{\mcitedefaultseppunct}\relax
\EndOfBibitem
\bibitem[Balzer and Stock(2005)Balzer, and Stock]{balzer2005}
Balzer,~B.; Stock,~G. Modeling of Decoherence and Dissipation in Nonadiabatic
  Photoreactions by an Effective-Scaling Nonsecular Redfield Algorithm.
  \emph{Chem. Phys.} \textbf{2005}, \emph{310}, 33--41\relax
\mciteBstWouldAddEndPuncttrue
\mciteSetBstMidEndSepPunct{\mcitedefaultmidpunct}
{\mcitedefaultendpunct}{\mcitedefaultseppunct}\relax
\EndOfBibitem
\bibitem[Gonz{\'a}lez-Luque \latin{et~al.}(2000)Gonz{\'a}lez-Luque, Garavelli,
  Bernardi, Merch{\'a}n, Robb, and Olivucci]{Olivucci2000PNAS}
Gonz{\'a}lez-Luque,~R.; Garavelli,~M.; Bernardi,~F.; Merch{\'a}n,~M.;
  Robb,~M.~A.; Olivucci,~M. Computational Evidence in Favor of a Two-State,
  Two-Mode Model of the Retinal Chromophore Photoisomerization. \emph{Proc.
  Nat. Acad. Sci.} \textbf{2000}, \emph{97}, 9379--9384\relax
\mciteBstWouldAddEndPuncttrue
\mciteSetBstMidEndSepPunct{\mcitedefaultmidpunct}
{\mcitedefaultendpunct}{\mcitedefaultseppunct}\relax
\EndOfBibitem
\bibitem[Nakamichi and Okada(2006)Nakamichi, and Okada]{NakamichiOkada2006ACIE}
Nakamichi,~H.; Okada,~T. Crystallographic Analysis of Primary Visual
  Photochemistry. \emph{Angew. Chem. Int.} \textbf{2006}, \emph{45},
  4270--4273\relax
\mciteBstWouldAddEndPuncttrue
\mciteSetBstMidEndSepPunct{\mcitedefaultmidpunct}
{\mcitedefaultendpunct}{\mcitedefaultseppunct}\relax
\EndOfBibitem
\bibitem[Tscherbul and Brumer(2014)Tscherbul, and Brumer]{tscherbul2014}
Tscherbul,~T.~V.; Brumer,~P. Excitation of Biomolecules with Incoherent Light:
  Quantum Yield for the Photoisomerization of Model Retinal. \emph{J. Phys.
  Chem. A} \textbf{2014}, \emph{118}, 3100--3111\relax
\mciteBstWouldAddEndPuncttrue
\mciteSetBstMidEndSepPunct{\mcitedefaultmidpunct}
{\mcitedefaultendpunct}{\mcitedefaultseppunct}\relax
\EndOfBibitem
\bibitem[Dodin \latin{et~al.}(2018)Dodin, Tscherbul, Alicki, Vutha, and
  Brumer]{Dodin2018PRA}
Dodin,~A.; Tscherbul,~T.; Alicki,~R.; Vutha,~A.; Brumer,~P. Secular versus
  Nonsecular Redfield Dynamics and Fano Coherences in Incoherent Excitation: An
  Experimental Proposal. \emph{Phys. Rev. A} \textbf{2018}, \emph{97},
  013421\relax
\mciteBstWouldAddEndPuncttrue
\mciteSetBstMidEndSepPunct{\mcitedefaultmidpunct}
{\mcitedefaultendpunct}{\mcitedefaultseppunct}\relax
\EndOfBibitem
\bibitem[Sala and Egorova(2018)Sala, and Egorova]{sala2018}
Sala,~M.; Egorova,~D. Quantum Dynamics of Multi-Dimensional Rhodopsin
  Photoisomerization Models: Approximate versus Accurate Treatment of the
  Secondary Modes. \emph{Chem. Phys.} \textbf{2018}, \emph{515}, 164--176\relax
\mciteBstWouldAddEndPuncttrue
\mciteSetBstMidEndSepPunct{\mcitedefaultmidpunct}
{\mcitedefaultendpunct}{\mcitedefaultseppunct}\relax
\EndOfBibitem
\bibitem[Hahn and Stock(2000)Hahn, and Stock]{hahn2000CP}
Hahn,~S.; Stock,~G. Femtosecond Secondary Emission Arising from the
  Nonadiabatic Photoisomerization in Rhodopsin. \emph{Chem. Phys.}
  \textbf{2000}, \emph{259}, 297--312\relax
\mciteBstWouldAddEndPuncttrue
\mciteSetBstMidEndSepPunct{\mcitedefaultmidpunct}
{\mcitedefaultendpunct}{\mcitedefaultseppunct}\relax
\EndOfBibitem
\bibitem[Prokhorenko \latin{et~al.}(2006)Prokhorenko, Nagy, Waschuk, Brown,
  Birge, and Miller]{prokhorenko2006}
Prokhorenko,~V.~I.; Nagy,~A.~M.; Waschuk,~S.~A.; Brown,~L.~S.; Birge,~R.~R.;
  Miller,~R.~D. Coherent Control of Retinal Isomerization in Bacteriorhodopsin.
  \emph{Science} \textbf{2006}, \emph{313}, 1257--1261\relax
\mciteBstWouldAddEndPuncttrue
\mciteSetBstMidEndSepPunct{\mcitedefaultmidpunct}
{\mcitedefaultendpunct}{\mcitedefaultseppunct}\relax
\EndOfBibitem
\bibitem[Fujisaki and Takatsuka(2001)Fujisaki, and Takatsuka]{Takatsuka2001PRE}
Fujisaki,~H.; Takatsuka,~K. Chaos Induced by Quantum Effect due to Breakdown of
  the Born-Oppenheimer Adiabaticity. \emph{Phys. Rev. E} \textbf{2001},
  \emph{63}, 066221\relax
\mciteBstWouldAddEndPuncttrue
\mciteSetBstMidEndSepPunct{\mcitedefaultmidpunct}
{\mcitedefaultendpunct}{\mcitedefaultseppunct}\relax
\EndOfBibitem
\bibitem[Marsili \latin{et~al.}(2020)Marsili, Olivucci, Lauvergnat, and
  Agostini]{Marsili2020JCTC}
Marsili,~E.; Olivucci,~M.; Lauvergnat,~D.; Agostini,~F. Quantum and
  Quantum-Classical Studies of the Photoisomerization of a Retinal Chromophore
  Model. \emph{J. Chem. Theo. Comp.} \textbf{2020}, \emph{16}, 6032--6048\relax
\mciteBstWouldAddEndPuncttrue
\mciteSetBstMidEndSepPunct{\mcitedefaultmidpunct}
{\mcitedefaultendpunct}{\mcitedefaultseppunct}\relax
\EndOfBibitem
\bibitem[Heller(1990)]{Heller1990}
Heller,~E. Mode Mixing and Chaos Induced by Potential Surface Crossings.
  \emph{J. Chem. Phys.} \textbf{1990}, \emph{92}, 1718\relax
\mciteBstWouldAddEndPuncttrue
\mciteSetBstMidEndSepPunct{\mcitedefaultmidpunct}
{\mcitedefaultendpunct}{\mcitedefaultseppunct}\relax
\EndOfBibitem
\bibitem[Leitner \latin{et~al.}({1996})Leitner, K{\"{o}}ppel, and
  Cederbaum]{LeitnerCederbaum1996}
Leitner,~D.~M.; K{\"{o}}ppel,~H.; Cederbaum,~L.~S. {Statistical Properties of
  Molecular Spectra and Molecular Dynamics: Analysis of Their Correspondence in
  NO2 and C2H4}. \emph{{J. Chem. Phys.}} \textbf{{1996}}, \emph{{104}},
  {434}\relax
\mciteBstWouldAddEndPuncttrue
\mciteSetBstMidEndSepPunct{\mcitedefaultmidpunct}
{\mcitedefaultendpunct}{\mcitedefaultseppunct}\relax
\EndOfBibitem
\bibitem[Haller \latin{et~al.}(1983)Haller, K{\"{o}}ppel, and
  Cederbaum]{Cederbaum1983}
Haller,~E.; K{\"{o}}ppel,~H.; Cederbaum,~L.~S. On the Statistical Behaviour of
  Molecular Vibronic Energy Levels. \emph{Chem. Phys. Lett.} \textbf{1983},
  \emph{101}, 215\relax
\mciteBstWouldAddEndPuncttrue
\mciteSetBstMidEndSepPunct{\mcitedefaultmidpunct}
{\mcitedefaultendpunct}{\mcitedefaultseppunct}\relax
\EndOfBibitem
\bibitem[Santoro \latin{et~al.}({2007})Santoro, Lami, and
  Olivucci]{SantoroOlivucci2007}
Santoro,~F.; Lami,~A.; Olivucci,~M. {Complex Excited Dynamics around a Plateau
  on a Retinal-like Potential Surface: Chaos, Multi-Exponential Decays and
  Quantum/Classical Differences}. \emph{{Theo. Chem. Acct.}} \textbf{{2007}},
  \emph{{117}}, {1061--1072}\relax
\mciteBstWouldAddEndPuncttrue
\mciteSetBstMidEndSepPunct{\mcitedefaultmidpunct}
{\mcitedefaultendpunct}{\mcitedefaultseppunct}\relax
\EndOfBibitem
\bibitem[Biggs(1993)]{BiggsGraphTheoryBook}
Biggs,~N. \emph{Algebraic Graph Theory}; Cambridge University Press, 1993\relax
\mciteBstWouldAddEndPuncttrue
\mciteSetBstMidEndSepPunct{\mcitedefaultmidpunct}
{\mcitedefaultendpunct}{\mcitedefaultseppunct}\relax
\EndOfBibitem
\bibitem[Schnedermann \latin{et~al.}(2018)Schnedermann, Yang, Liebel, Spillane,
  Lugtenburg, Fern{\'a}ndez, Valentini, Schapiro, Olivucci, Kukura, and
  Mathies]{schnedermann2018}
Schnedermann,~C.; Yang,~X.; Liebel,~M.; Spillane,~K.; Lugtenburg,~J.;
  Fern{\'a}ndez,~I.; Valentini,~A.; Schapiro,~I.; Olivucci,~M.; Kukura,~P.
  \latin{et~al.}  Evidence for a Vibrational Phase-Dependent Isotope Effect on
  the Photochemistry of Vision. \emph{Nat. Chem.} \textbf{2018}, \emph{10},
  449\relax
\mciteBstWouldAddEndPuncttrue
\mciteSetBstMidEndSepPunct{\mcitedefaultmidpunct}
{\mcitedefaultendpunct}{\mcitedefaultseppunct}\relax
\EndOfBibitem
\bibitem[Kim \latin{et~al.}(2001)Kim, Tauber, and Mathies]{kim2001}
Kim,~J.~E.; Tauber,~M.~J.; Mathies,~R.~A. Wavelength Dependent cis-trans
  Isomerization in Vision. \emph{Biochem.} \textbf{2001}, \emph{40},
  13774--13778\relax
\mciteBstWouldAddEndPuncttrue
\mciteSetBstMidEndSepPunct{\mcitedefaultmidpunct}
{\mcitedefaultendpunct}{\mcitedefaultseppunct}\relax
\EndOfBibitem
\bibitem[Sovdat \latin{et~al.}(2012)Sovdat, Bassolino, Liebel, Schnedermann,
  Fletcher, and Kukura]{Kukura2012JACS}
Sovdat,~T.; Bassolino,~G.; Liebel,~M.; Schnedermann,~C.; Fletcher,~S.~P.;
  Kukura,~P. Backbone Modification of Retinal Induces Protein-like Excited
  State Dynamics in Solution. \emph{J. Amer. Chem. Soc.} \textbf{2012},
  \emph{134}, 8318--8320\relax
\mciteBstWouldAddEndPuncttrue
\mciteSetBstMidEndSepPunct{\mcitedefaultmidpunct}
{\mcitedefaultendpunct}{\mcitedefaultseppunct}\relax
\EndOfBibitem
\bibitem[Grobe \latin{et~al.}(1988)Grobe, Haake, and
  Sommers]{GrobeHaakeSommers1988PRL}
Grobe,~R.; Haake,~F.; Sommers,~H.-J. Quantum Distinction of Regular and Chaotic
  Dissipative Motion. \emph{Phys. Rev. Lett.} \textbf{1988}, \emph{61},
  1899--1902\relax
\mciteBstWouldAddEndPuncttrue
\mciteSetBstMidEndSepPunct{\mcitedefaultmidpunct}
{\mcitedefaultendpunct}{\mcitedefaultseppunct}\relax
\EndOfBibitem
\bibitem[Akemann \latin{et~al.}(2019)Akemann, Kieburg, Mielke, and
  Prosen]{Prosen2019PRL}
Akemann,~G.; Kieburg,~M.; Mielke,~A.; Prosen,~T. Universal Signature from
  Integrability to Chaos in Dissipative Open Quantum Systems. \emph{Phys. Rev.
  Lett.} \textbf{2019}, \emph{123}, 254101\relax
\mciteBstWouldAddEndPuncttrue
\mciteSetBstMidEndSepPunct{\mcitedefaultmidpunct}
{\mcitedefaultendpunct}{\mcitedefaultseppunct}\relax
\EndOfBibitem
\bibitem[S\'a \latin{et~al.}(2020)S\'a, Ribeiro, and Prosen]{Prosen2020PRX}
S\'a,~L.; Ribeiro,~P.; Prosen,~T. Complex Spacing Ratios: A Signature of
  Dissipative Quantum Chaos. \emph{Phys. Rev. X} \textbf{2020}, \emph{10},
  021019\relax
\mciteBstWouldAddEndPuncttrue
\mciteSetBstMidEndSepPunct{\mcitedefaultmidpunct}
{\mcitedefaultendpunct}{\mcitedefaultseppunct}\relax
\EndOfBibitem
\bibitem[Jalabert and Pastawski(2001)Jalabert, and
  Pastawski]{JalabertPastawski2001PRL}
Jalabert,~R.~A.; Pastawski,~H.~M. Environment-Independent Decoherence Rate in
  Classically Chaotic Systems. \emph{Phys. Rev. Lett.} \textbf{2001},
  \emph{86}, 2490--2493\relax
\mciteBstWouldAddEndPuncttrue
\mciteSetBstMidEndSepPunct{\mcitedefaultmidpunct}
{\mcitedefaultendpunct}{\mcitedefaultseppunct}\relax
\EndOfBibitem
\bibitem[Cucchietti \latin{et~al.}(2003)Cucchietti, Dalvit, Paz, and
  Zurek]{CucchiettiZurek2003PRL}
Cucchietti,~F.~M.; Dalvit,~D. A.~R.; Paz,~J.~P.; Zurek,~W.~H. Decoherence and
  the Loschmidt Echo. \emph{Phys. Rev. Lett.} \textbf{2003}, \emph{91},
  210403\relax
\mciteBstWouldAddEndPuncttrue
\mciteSetBstMidEndSepPunct{\mcitedefaultmidpunct}
{\mcitedefaultendpunct}{\mcitedefaultseppunct}\relax
\EndOfBibitem
\bibitem[Chen \latin{et~al.}(2016)Chen, Gelin, Chernyak, Domcke, and
  Zhao]{DomckeZhao2016}
Chen,~L.; Gelin,~M.~F.; Chernyak,~V.~Y.; Domcke,~W.; Zhao,~Y. Dissipative
  Dynamics at Conical Intersections: Simulations with the Hierarchy Equations
  of Motion Method. \emph{Faraday Discuss.} \textbf{2016}, \emph{194},
  61--80\relax
\mciteBstWouldAddEndPuncttrue
\mciteSetBstMidEndSepPunct{\mcitedefaultmidpunct}
{\mcitedefaultendpunct}{\mcitedefaultseppunct}\relax
\EndOfBibitem
\end{mcitethebibliography}
\end{document}